\documentclass[11pt,twoside]{article}
\usepackage{./asp2014}
\usepackage{graphicx}

\aspSuppressVolSlug
\resetcounters

\begin{document}

\title{Using an Artificial Neural Network to Classify Multi-component Emission Line Fits}
\author{Elise J~Hampton,$^1$ Brent~Groves,$^1$ Anne~Medling,$^1$ Rebecca~Davies,$^1$ Mike~Dopita,$^1$ I-Ting~Ho,$^2$ Melanie~Kaasinen,$^1$ Lisa~Kewley,$^1$ Sarah~Leslie,$^1$ Rob~Sharp,$^1$ Sarah M~Sweet,$^1$ Adam D~Thomas,$^1$ SAMI Survey Team$^3$, and S7 Team$^4$  
\affil{$^1$Research School of Astronomy \& Astrophysics, Australian National University, Canberra, ACT 2611, Australia; \email{contact: elise.hampton@anu.edu.au}}
\affil{$^2$Institute for Astronomy, University of Hawaii, 2680 Woodlawn Drive, Honolulu, HI 96822, USA}
\affil{$^3$SAMI Galaxy Survey team members and status: \url{http://sami-survey.org/members}}
\affil{$^4$S7 team members and status: \url{http://miocene.anu.edu.au/S7/}}}

\paperauthor{Elise J~Hampton}{elise.hampton@anu.edu.au}{}{Australian National University}{Research School of Astronomy \& Astrophysics}{Canberra}{ACT}{2611}{Australia}
\paperauthor{Brent~Groves}{brent.groves@anu.edu.au}{}{Australian National University}{Research School of Astronomy \& Astrophysics}{Canberra}{ACT}{2611}{Australia}
\paperauthor{Anne~Medling}{anne.medling@anu.edu.au}{}{Australian National University}{Research School of Astronomy \& Astrophysics}{Canberra}{ACT}{2611}{Australia}
\paperauthor{Rebecca~Davies}{rebecca.davies@anu.edu.au}{}{Australian National University}{Research School of Astronomy \& Astrophysics}{Canberra}{ACT}{2611}{Australia}
\paperauthor{Mike~Dopita}{michael.dopita@anu.edu.au}{}{Australian National University}{Research School of Astronomy \& Astrophysics}{Canberra}{ACT}{2611}{Australia}
\paperauthor{I-Ting~Ho}{itho@Ifa.hawaii.edu}{}{University of Hawaii}{Institute for Astronomy}{Honolulu}{HI}{96822}{USA}
\paperauthor{Melanie~Kaasinen}{melanie.kaasinen@anu.edu.au}{}{Australian National University}{Research School of Astronomy \& Astrophysics}{Canberra}{ACT}{2611}{Australia}
\paperauthor{Lisa~Kewley}{lisa.kewley@anu.edu.au}{}{Australian National University}{Research School of Astronomy \& Astrophysics}{Canberra}{ACT}{2611}{Australia}
\paperauthor{Sarah~Leslie}{sarah.k.leslie@gmail.com}{}{Australian National University}{Research School of Astronomy \& Astrophysics}{Canberra}{ACT}{2611}{Australia}
\paperauthor{Rob~Sharp}{rob.sharp@anu.edu.au}{}{Australian National University}{Research School of Astronomy \& Astrophysics}{Canberra}{ACT}{2611}{Australia}
\paperauthor{Sarah M~Sweet}{sarah@sarahsweet.com.au}{}{Australian National University}{Research School of Astronomy \& Astrophysics}{Canberra}{ACT}{2611}{Australia}
\paperauthor{Adam D~Thomas}{adam.thomas@anu.edu.au}{}{Australian National University}{Research School of Astronomy \& Astrophysics}{Canberra}{ACT}{2611}{Australia}

\begin{abstract}
We present \texttt{The Machine}, an artificial neural network (ANN) capable of differentiating between the numbers of Gaussian components needed to describe the emission lines of Integral Field Spectroscopic (IFS) observations. Here we show the preliminary results of the S7 first data release \citep[Siding Spring Southern Seyfert Spectroscopic Snapshot Survey,][]{Dopita:2015aa} and SAMI Galaxy Survey \citep[Sydney-AAO Multi-object Integral Field Unit,][]{Croom:2012aa} to classify whether the emission lines in each spatial pixel are composed of 1, 2, or 3 different Gaussian components. Previously this classification has been done by individual people, taking an hour per galaxy. This time investment is no longer feasible with the large spectroscopic surveys coming online. 
\end{abstract}

\section{Introduction}

Integral Field Spectroscopy (IFS) is changing our approach to studying galaxy evolution. Surveys such as CALIFA \citep[Calar Alto Legacy Integral Field Area,][]{Sanchez:2012aa}, SAMI Galaxy Survey \citep[Sydney-AAO Multi-object Integral Field Unit,][]{Croom:2012aa}, MaNGA \citep[Mapping Nearby Galaxies at Apache Point Observatory,][]{Bundy:2015aa}, and S7 \citep[Siding Spring Southern Seyfert Spectroscopic Snapshot Survey,][]{Dopita:2014aa} are building databases of hundreds to thousands of galaxies to explore galaxy evolution as a function of morphological and spectroscopic classification.

\begin{figure}[ht!]
\plotone[height=4cm]{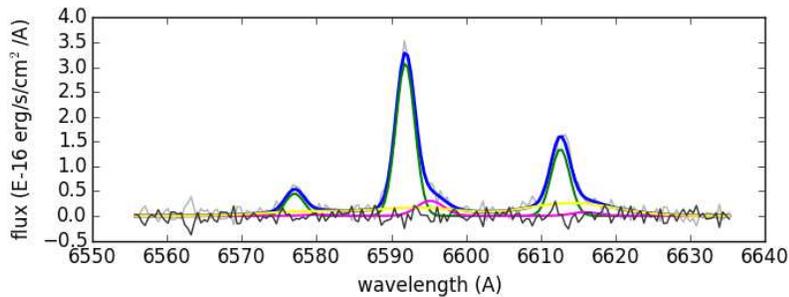}
\caption{An example of 3 Gaussian components being fit to the [NII] and H$\alpha$ emission lines. Grey is the data, coloured are the Gaussian components fit, and black is the residual from subtracting the fits from the data.}
\label{fig1}
\end{figure}

This type of data comes at a price: data volume. Data reduction through pipelines \citep[e.g.][]{Husemann:2013aa,Sharp:2015aa,Allen:2015aa} addresses the preparation of raw data to an analysis stage, but understanding the contents of these data cubes remains a significant challenge. Automated continuum and absorption line fitting is routinely used to understand the stellar populations within galaxies. Emission line fitting provides insight into active star formation, AGN (active galactic nuclei) activity, and shock properties of galaxies. This type of pre-analysis is time consuming for integral field unit (IFU) surveys and no longer feasibly done by hand. We now understand that there can also be multiple physical processes present, creating further steps to our pre-analysis. Automated emission line fitting, including the fitting of multiple components (Fig. \ref{fig1}), is currently in use \citep[e.g. LZIFU,][]{Ho:press} but still requires human input to decide the best number of components to describe the spectra. We aim to remove this time consuming human input and streamline multi-component emission line fitting for large surveys using a machine learning algorithm: an artificial neural network. 

\section{The Machine}
The use of machine learning in Astronomy is not a new idea. Recent examples include the prediction of solar flares \citep[e.g.][]{Bobra:2015aa}, understanding Gamma-ray emission from AGNs \citep[e.g.][]{Doert:2014aa,Hassan:2013aa}, and the classification of galaxy types \citep[e.g.][]{Kuminski:2014aa}. Machine learning covers a wide range of distinct classes of artificial intelligence (AI) such as Artificial Neural Networks (ANNs), Support Vector Machines (SVMs), and Random Forest algorithms, that learn without being explicitly programmed. Each have their benefits and weaknesses but all are based on the same underlying principle, they learn from a training set and create models to be used to predict outcomes. We have chosen to use an Artificial Neural Network (ANN) to build a classification model for multi-component emission line fitting.

An Artificial Neural Network (ANN) is a computer system comprised of nodes, or units, which perform calculations with a pre-determined equation. In our case we use a sigmoid function $1/(1+exp(x))$. These nodes sit in layers which have different jobs depending on where they sit in the ANN design. 
Our ANN\footnote{For an in depth description of how an ANN, and specifically our ANN \texttt{The Machine}, works see \citet{Hampton:2015}}, which we have called \texttt{The Machine}\footnote{The name of our ANN has been based on the Artificial Intelligence built by a Mr Finch in the TV series 'Person of Interest'. The outer program that controls the input and output of \texttt{The Machine} is called \texttt{Finch}.}, has two hidden layers with 15 nodes in each layer. The input layer has 86 input parameters to describe the emission-line fits, making up the feature vector for each example and the output layer has 3 or 4 nodes corresponding to the different classifications we are looking for: 0, 1, 2, 3 components or 1, 2, 3 components. During training we give \texttt{The Machine} examples of classifications we expect it to see. 

Here, we present the preliminary results of applying \texttt{The Machine} to the S7 first data release \citep{Dopita:2015aa} and to the current SAMI survey galaxies \citep{Croom:2012aa} to classify whether the emission lines in each spatial pixel are composed of 1, 2, or 3 different components (or no emission lines, classed as 0, for S7).

The training sets are produced by two separate teams of trainers, 3 spectroscopists for S7 and 5 for SAMI. Each trainer classifies the spaxels (spatial pixels) of nine galaxies in their survey with their recommendation of the best number of components. Then each galaxy is given a single component map corresponding to the examples where at least 3 trainers agreed on the classification. Classifications not agreed on by at least 3 trainers are excluded from the training.

The overall accuracy of \texttt{The Machine} is described as the percentage of classifications \texttt{The Machine} identified as being the same as our combined trainers. The accuracy can be split into the recall and precision \texttt{The Machine} obtains for each individual classification so as to understand where it may need more training. 

The recall and precision values are calculated for each classification \texttt{The Machine} can make. Recall measures the consistency \texttt{The Machine} has for each classification related to how often it misclassifies an example of that component number. For example, If \texttt{The Machine} correctly classifies 200 examples as 1-components but misclassifies 50 1-component examples as 2- or 3-components it has a recall of 0.8 or 80\% recall for 1-component classifications. A recall value is calculated for each classification, 0 to 3. 

Precision measures how often \texttt{The Machine} will misclassify an example as a particular number of components. For example, If \texttt{The Machine} correctly classifies 200 examples as 1-components but also classifies 25 2- and 3-component examples as 1-components it has a precision value of 0.89 or 89\% precision in 1-component classifications. Precision values are, just like with recall values, calculated for each classification, 0 to 3.   

\begin{figure}
\plotone[height=4.5cm]{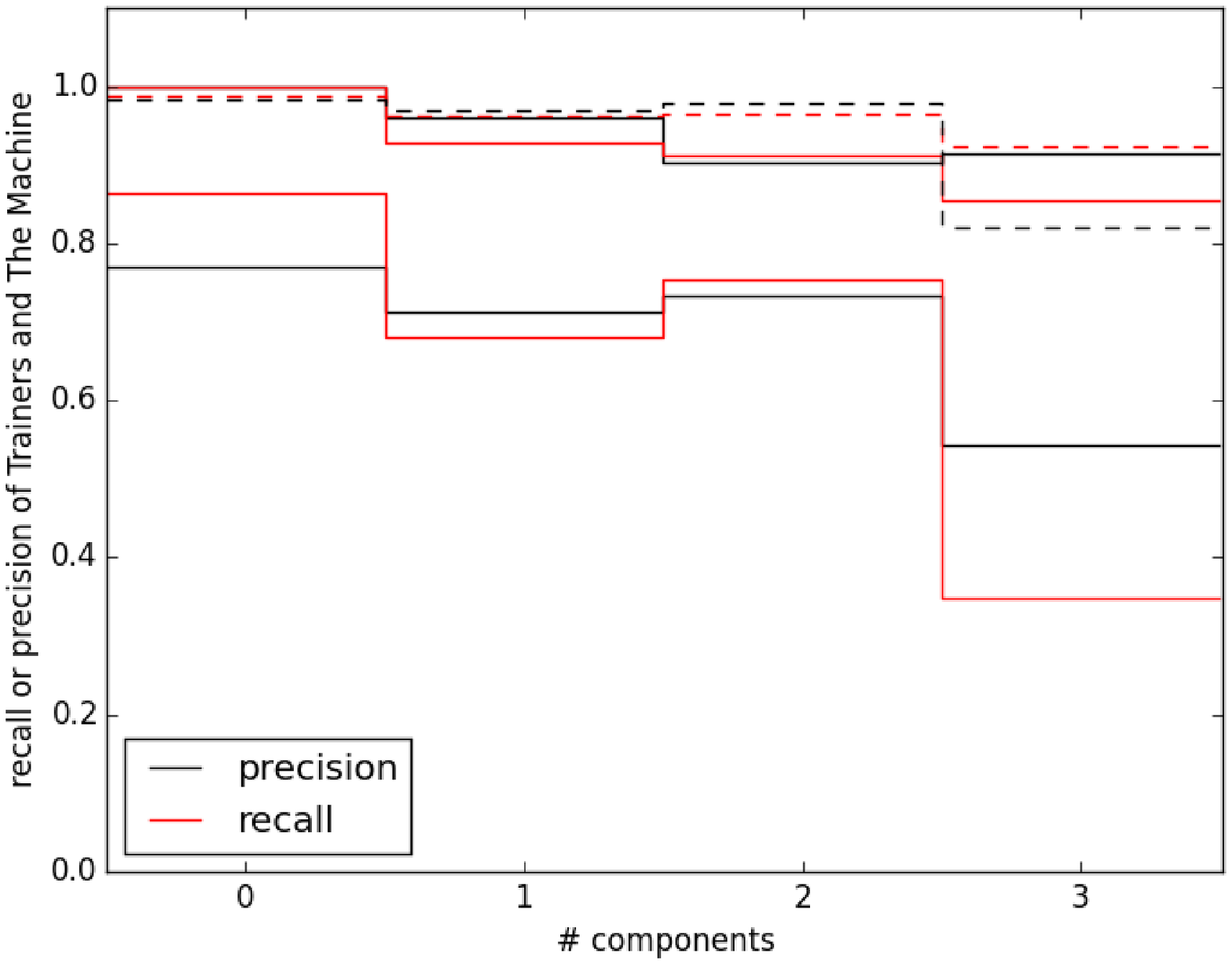}
\plotone[height=4.5cm]{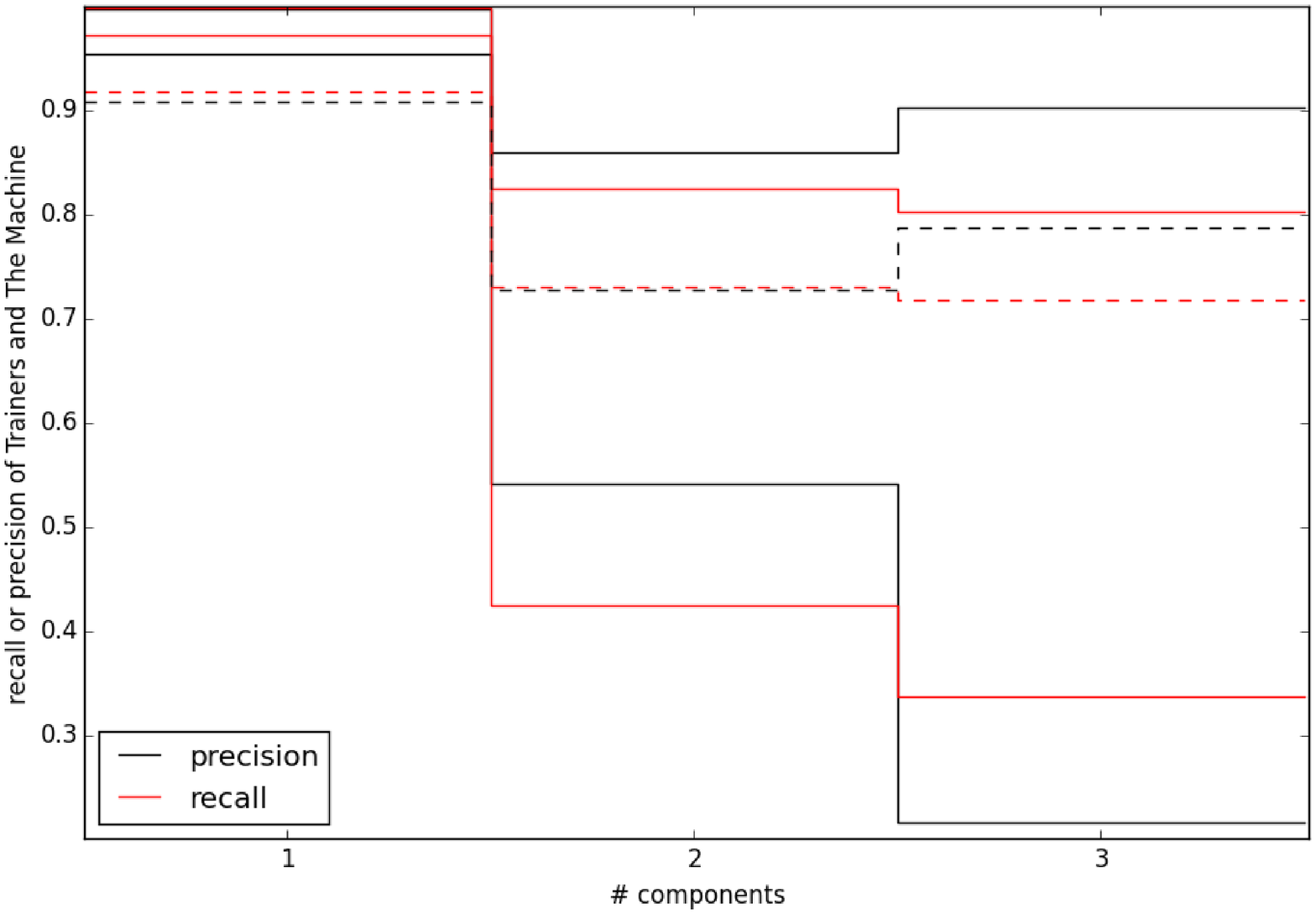}
\caption{We present the minimum and maximum values of the precision and recall from our S7 (left) and SAMI (right) trainers in solid lines. The dashed lines show \texttt{The Machine's} results from training. Figures from \citet{Hampton:2015}}
\label{fig4}
\end{figure}

To understand how \texttt{The Machine's} recall and precision compares to our trainers we have compared our trainers to each other. The recall and precision of our Trainers to each other are presented in Fig. \ref{fig4} as the solid lines. The dashed lines are the recall and precision values obtained by \texttt{The Machine} after training. \texttt{The Machine} can be seen to have a higher recall than any of our trainers, and a consistently high precision across all components in S7 galaxies. With SAMI data we see that \texttt{The Machine} has less capability for recall and precision on 1-component fits but similar recall and precision for 2 and 3 component examples.

The classifications of individual galaxies by our Trainers takes 1 hour per galaxy. People still make good trainers or classifiers for small surveys. For a survey like SAMI ($\sim$ 3000 galaxies), this is 125 straight days of classifying. \texttt{The Machine} is capable of classifying every spaxel in over a thousand galaxies, including creating component maps, in $\sim$8 minutes \citep{Hampton:2015}.

\section{Conclusion}
Complex emission line fitting of IFU data is not new. With more surveys nowadays observing hundreds to thousands of galaxies with IFS observations, automated complex emission line fitting is a must. LZIFU has automated the fitting process for up to 3 Gaussian components but does not have the capacity to tell us which of its set of fits best describe the spectra for that particular spaxel. Our ANN, \texttt{The Machine}, does have that capacity; indicating that the complexities of differentiating between multi-component fits can be solved reliably and rapidly using machine learning. 

We have built \texttt{The Machine} to take in information produced by LZIFU and output the best fit classification to each individual spaxel in each galaxy of a survey. It is a fast, self-consistent, and reliable system that replaces the need for years of manual work by astronomers. The breakdown of the accuracy  into recall and precision of \texttt{The Machine} shows that it is indistinguishable from human trainers. 

Our research has been to create a way to easily determine the best fit results of LZIFU for large surveys. We have shown that it is possible to use machine learning to create a fast and self-consistent classification system. This paper describes the preliminary results of the S7 and SAMI surveys. For a full description of \texttt{The Machine} and our full analysis of the training see \citet{Hampton:2015}.

\bibliography{O8-3.bib}

\end{document}